\documentclass[twoside,11pt]{7ssmmp} 

\usepackage{amsfonts}
\usepackage{amssymb}
\usepackage{amsmath}
\usepackage{fancyhdr}    
\usepackage{equations}
\usepackage{epic}
\usepackage{eepic}
\usepackage{epsf}
\usepackage{graphicx}
\usepackage{rotating}

\newcommand\rf[1]{(\ref{eq:#1})}
\newcommand\lab[1]{\label{eq:#1}}
\newcommand\nonu{\nonumber}
\newcommand\br{\begin{eqnarray}}
\newcommand\er{\end{eqnarray}}
\newcommand\be{\begin{equation}}
\newcommand\ee{\end{equation}}

\newcommand\lb{\lbrack}
\newcommand\rb{\rbrack}

\newcommand\llb{\left\lbrack}
\newcommand\rrb{\right\rbrack}

\renewcommand\({\left(}
\renewcommand\){\right)}
\newcommand\bv{\bigm\vert}               

\newcommand\bc{\begin{center}}
\newcommand\ec{\end{center}}

\newcommand\partder[2]{\frac{{\partial {#1}}}{{\partial {#2}}}}

\renewcommand\a{\alpha}
\renewcommand\b{\beta}

\renewcommand\d{\delta}

\newcommand\vareps{\varepsilon}
\newcommand\g{\gamma}
\newcommand\G{\Gamma}

\newcommand\h{\frac{1}{2}}
\renewcommand\k{\kappa}
\renewcommand\l{\lambda}
\renewcommand\L{\Lambda}
\newcommand\m{\mu}
\newcommand\n{\nu}

\newcommand\vp{\varphi}
\renewcommand\P{\Phi}
\newcommand\pa{\partial}

\newcommand\pr{\prime}

\newcommand\s{\sigma}

\renewcommand\t{\tau}
\renewcommand\th{\theta}

\newcommand\wti{\widetilde}

\newcommand\cA{{\mathcal A}}

\newcommand\cM{{\mathcal M}}

\newcommand{\ct}[1]{\cite{#1}}
\newcommand{\bib}[1]{\bibitem{#1}}

\newcommand\PRL[3]{\textsl{Phys. Rev. Lett.} \textbf{#1} (#2) #3}
\newcommand\NPB[3]{\textsl{Nucl. Phys.} \textbf{B#1} (#2) #3}

\newcommand\PRD[3]{\textsl{Phys. Rev.} \textbf{D#1} (#2) #3}

\newcommand\PLB[3]{\textsl{Phys. Lett.} \textbf{#1B} (#2) #3}
\newcommand\CQG[3]{\textsl{Class. Quantum Grav.} \textbf{#1} (#2) #3}

\newcommand\AoP[3]{\textsl{Ann. of Phys.} \textbf{#1} (#2) #3}

\newcommand\IJMPA[3]{\textsl{Int. J. Mod. Phys.} \textbf{A#1} (#2) #3}

\newcommand\MPLA[3]{\textsl{Mod. Phys. Lett.} \textbf{A#1} (#2) #3}

\renewcommand{\headrulewidth}{0.4pt}

\begin{document}
 \baselineskip=11pt

\title{Gravity, Nonlinear Gauge Fields and Charge Confinement/Deconfinement
\hspace{.25mm}\thanks{\,Work supported in
part by Bulgarian National Science Foundation grant DO 02-257}}
\author{\bf{Eduardo Guendelman and Alexander Kaganovich}\hspace{.25mm}\thanks{\,e-mail address:
guendel@bgu.ac.il, alexk@bgu.ac.il}
\\ \normalsize{Physics Department, Ben Gurion University of the Negev} \\
\normalsize{Beer Sheva, Israel} \vspace{2mm} \\ 
\bf{Emil Nissimov and Svetlana Pacheva}\hspace{.25mm}\thanks{\,e-mail
address: nissimov@inrne.bas.bg, svetlana@inrne.bas.bg}
\\ \normalsize{Institute for Nuclear Research and Nuclear Energy}\\
\normalsize{Bulgarian Academy of Sciences, Sofia, Bulgaria} }

\date{}

\maketitle

\begin{abstract}
We discuss in some detail the properties of  gravity (including $f(R)$-gravity)
coupled to non-standard nonlinear gauge field system containing a square root of 
the usual Maxwell Lagrangian $-\frac{f_0}{2}\sqrt{-F^2}$. The latter is known
to produce in flat spacetime a QCD-like confinement. Inclusion of gravity
triggers various physically interesting effects: new mechanism for dynamical 
generation of cosmological constant; non-standard black hole solutions with
constant vacuum electric field and with ``hedge-hog''-type spacetime
asymptotics, which are shown to obey the first law of black hole thermodynamics; 
new ``tubelike'' solutions of Levi-Civita-Bertotti-Robinson type;
charge-''hiding'' and charge-confining ``thin-shell'' wormhole solutions;
dynamical effective gauge couplings and confinement-deconfinement transition effect
when coupled to quadratic $R^2$-gravity.
\end{abstract}

\section{Introduction}

We consider gravity, including $f(R)$-gravity \ct{f(R)-grav}, coupled to 
{\em non-standard} nonlinear gauge field system containing a square root of the 
ordinary Maxwell Lagrangian $-\frac{f_0}{2}\sqrt{-F^2}$. In flat spacetime the 
latter model has been shown \ct{GG} to produce a {\em QCD-like confinement}.

We exhibit several interesting features of the above system (see also 
Refs.\ct{gravcornell-f(R)grav,hiding-hideconfine}) :
\begin{itemize}
\item
New mechanism for {\em dynamical} generation of cosmological constant
due to nonlinear gauge field dynamics: $\L_{\rm eff} = \L_0 + 2\pi f_0^2$ 
~($\L_0$ -- bare cosmological constant, may be absent at all).
\item
Non-standard black hole solutions of Reissner-Nordstr{\"o}m-\-(anti-)\-de-Sitter
type containing a {\em constant radial vacuum electric field} (in addition to the
Coulomb one), in particular, in electrically neutral black holes of
Schwarzschild-(anti-)de-Sitter type. It is shown that these non-standard
black holes obey the first law of black hole thermodynamics.
\item
In case of vanishing effective cosmological constant $\L_{\rm eff}$
(\textsl{i.e.}, $\L_0 < 0\; ,\; |\L_0|=2\pi f_0^2$) the resulting
Reissner-Nordstr{\"o}m-type black hole, apart from carrying an additional
constant vacuum electric field, turns out to be {\em non-asymptotically flat} -- 
a feature resembling the gravitational effect of a {\em hedgehog} \ct{hedgehog}.
\item
Appearance of {\em confining-type effective potential} in charged test particle 
dynamics in the above black hole backgrounds.
\item
New ``tubelike'' solutions of Levi-Civita-Bertotti-Robinson \ct{LC-BR} type, 
\textsl{i.e.}, with spacetime geometry of the form $\cM_2 \times S^2$, 
where $\cM_2$ is a two-dimensional anti-de Sitter, Rindler or de Sitter space 
depending on the relative strength of the electric field w.r.t. the coupling $f_0$
of the square-root gauge field term.
\end{itemize}

When in addition one or more {\em lightlike branes} are self-consistently
coupled to the above gravity/nonlinear-gauge-field system (as matter and
charge sources) they produce (``thin-shell'')  wormhole solutions displaying two 
novel physically interesting effects \ct{hiding-hideconfine}:
\begin{itemize}
\item
{\em ``Charge-hiding'' effect} - a genuinely charged matter source of gravity and 
electromagnetism may appear {\em electrically neutral} to an external observer 
-- a phenomenon opposite to the famous Misner-Wheeler ``charge without charge''
effect \ct{misner-wheeler};
\item
{\em Charge-confining ``tubelike'' wormhole} with two ``throats'' occupied by two
oppositely charged lightlike branes -- the whole electric flux is confined
within the finite-extent ``middle universe'' of generalized 
Levi-Civita-Bertotti-Robinson type 
-- no flux is escaping into the outer non-compact ``universes''. 
\end{itemize}

Additional interesting features appear when we couple the ``square-root''
confining nonlinear gauge field system to $f(R)$-gravity with $f(R)= R + \a R^2$ and
a dilaton. Reformulating the model in the physical ``Einstein'' frame we find
(cf. second Ref.\ct{gravcornell-f(R)grav}):
\begin{itemize}
\item
{\em ``Confinement-deconfinement'' transition} due to appearance of
``flat'' region in the effective dilaton potential;
\item
The effective gauge couplings as well as the induced
cosmological constant become {\em dynamical} depending on the dilaton v.e.v. 
In particular, a conventional Maxwell kinetic term for the gauge field is
{\em dynamically generated} even if absent in the original theory;
\item
{\em Regular black hole} solution ({\em no singularity} at $r=0$) 
{\em with confining vacuum electric field}:
the bulk spacetime consist of two regions -- an interior de Sitter and an exterior
Reissner-Nordstr\"{o}m-type (with ``hedgehog asymptotics'') glued together along 
their common horizon occupied by a charged lightlike brane. The latter also dynamically
determines the non-zero cosmological constant in the interior de-Sitter
region. This result is analogous to the regular black hole solution in the
case of ordinary Einstein gravity presented in Ref.\ct{Reg-BH} and will be
discussed in more detail in a subsequent paper.
\end{itemize}

Concluding the introductory remarks, let us briefly mention the principal
motivation for studying non-standard gauge field models with $\sqrt{-F^2}$.
G. `t Hooft has shown \ct{tHooft} that in any effective 
quantum gauge theory, which is able to describe linear confinement phenomena, 
the energy density of electrostatic field configurations should be a linear function
of the electric displacement field in the infrared region (the latter appearing as an
``infrared counterterm'').

The simplest way to realize these ideas in flat spacetime was proposed in
Refs.\ct{GG}:
\br
S = \int d^4 x L(F^2) \quad ,\quad
L(F^2) = -\frac{1}{4} F^2 - \frac{f_0}{2} \sqrt{-F^2} \; ,
\lab{GG-flat} \\
F^2 \equiv F_{\m\n} F^{\m\n} \quad ,\quad 
F_{\m\n} = \pa_\m A_\n - \pa_\n A_\m  \; ,
\nonu
\er
The square root of the Maxwell term naturally arises as a result of 
{\em spontaneous breakdown of scale symmetry} of 
the original scale-invariant Maxwell action with $f_0$ appearing as an integration 
constant responsible for the latter spontaneous breakdown.
For static field configurations the model \rf{GG-flat} yields an electric displacement
field $\vec{D} = \vec{E} - \frac{f_0}{\sqrt{2}}\frac{\vec{E}}{|\vec{E}|}$ and 
the corresponding  energy density turns out to be 
$\h \vec{E}^2 = \h |\vec{D}|^2 + \frac{f_0}{\sqrt{2}} |\vec{D}|+\frac{1}{4} f_0^2$, 
so that it indeed contains a term linear w.r.t. $|\vec{D}|$.
The model \rf{GG-flat} produces, when coupled to quantized fermions, a confining 
effective potential $V(r) = - \frac{\b}{r} + \g r$ (Coulomb plus linear one with 
$\g \sim f_0$) which is of the form of the well-known ``Cornell'' potential 
in the phenomenological description of quarkonium systems in QCD
\ct{cornell-potential}. 

\section{Einstein Gravity Coupled to Confining Nonlinear Gauge Field}
The pertinent action is given by ($R$-scalar curvature; $\L_0$ - 
bare cosmological constant, might be absent):
\br
S = \int d^4 x \sqrt{-G} \Bigl\lb \frac{R - 2\L_0}{16\pi} + L(F^2)\Bigr\rb 
\;\; ,\;\;
L(F^2) = - \frac{1}{4} F^2 - \frac{f_0}{2} \sqrt{- F^2} \; ,
\lab{gravity+GG} \\
F^2 \equiv F_{\k\l} F_{\m\n} G^{\k\m} G^{\l\n} \quad ,\quad 
F_{\m\n} = \pa_\m A_\n - \pa_\n A_\m \; .
\nonu
\er
\textbf{Remark.} One could start with the non-Abelian version of the 
gauge field action in \rf{gravity+GG}. Since we will be interested in static 
spherically symmetric solutions, the non-Abelian gauge theory effectively reduces 
to an Abelian one.

The corresponding equations of motion read accordingly -- Einstein equations:
\be
R_{\m\n} - \h G_{\m\n} R + \L_0 G_{\m\n} = 8\pi T^{(F)}_{\m\n} \; ,
\lab{einstein-eqs}
\ee
\be
T^{(F)}_{\m\n} = \Bigl( 1 - \frac{f_0}{\sqrt{-F^2}}\Bigr) F_{\m\k} F_{\n\l} G^{\k\l}
- \frac{1}{4} \Bigl( F^2 + 2f_0\sqrt{-F^2}\Bigr) G_{\m\n} \; ,
\lab{stress-tensor-F}
\ee
and nonlinear gauge field equations:
\be
\pa_\n \(\sqrt{-G}\Bigl( 1 - \frac{f_0}{\sqrt{-F^2}}\Bigr) F_{\k\l} G^{\m\k} G^{\n\l}\)=0
\; .
\lab{GG-eqs}
\ee
\textbf{Important remark}. Note the {\em non-zero} value of the trace of
energy-momentum tensor unlike ordinary Maxwell theory:
\br
T^{(F)} \equiv T^{(F)}_{\m\n} G^{\m\n} = - f_0 \sqrt{-F^2} \; .
\nonu
\er

Solving Eqs.\rf{einstein-eqs}--\rf{GG-eqs} we find new
\textsl{non-standard} Reissner-Nordstr{\"o}m-(anti-)de-Sitter-type black
holes depending on the sign of a dynamically generated cosmological constant 
$\L_{\mathrm{eff}}$:
\br
ds^2 = - A(r) dt^2 + \frac{dr^2}{A(r)} + r^2 \bigl(d\th^2 + \sin^2 \th d\vp^2\bigr)
\; ,
\lab{spherical-static} \\
A(r) = 1 - \sqrt{8\pi}|Q|f_0 - \frac{2m}{r} + \frac{Q^2}{r^2} 
- \frac{\L_{\mathrm{eff}}}{3} r^2 \quad ,\quad \L_{\mathrm{eff}} = 2\pi f_0^2 + \L_0 \; ,
\lab{CC-eff}
\er
with static spherically symmetric electric field containing apart from the Coulomb 
term an additional {\em constant} ``vacuum'' piece:
\be
F_{0r} = \frac{\vareps_F f_0}{\sqrt{2}} + \frac{Q}{\sqrt{4\pi}\, r^2} 
\quad ,\quad \vareps_F \equiv \mathrm{sign}(F_{0r}) = \mathrm{sign}(Q) \; .
\lab{cornell-sol}
\ee
The latter corresponds to a confining ``Cornell''-type \ct{cornell-potential} potential
$A_0 = -\frac{\vareps_F f_0}{\sqrt{2}}\, r + \frac{Q}{\sqrt{4\pi}\, r}$.
When $\L_{\mathrm{eff}}=0$, $A(r) \to 1 -\sqrt{8\pi}|Q|f_0$ for $r\to\infty$, 
\textsl{i.e.}, the black hole exhibits ``hedgehog'' \ct{hedgehog}
{\em non-flat-spacetime} asymptotics.

Furthermore, we find three distinct types of static solutions of ``tubelike''
Levi-Civita-Bertotti-Robinson \ct{LC-BR} type with spacetime geometry of the form 
$\cM_2 \times S^2$, where $\cM_2$ is some 2-dimensional manifold 
((anti-)de Sitter $(A)dS_2$, Rindler $Rind_2$):
\br
ds^2 = - A(\eta) dt^2 + \frac{d\eta^2}{A(\eta)} 
+ r_0^2 \bigl(d\th^2 + \sin^2 \th d\vp^2\bigr) \;\; ,\;\;  
-\infty < \eta <\infty \; ,
\lab{gen-BR-metric}\\
F_{0\eta} = c_F = \mathrm{const} \quad ,\quad 
\frac{1}{r_0^2} = 4\pi c_F^2 + \L_0 \; (=\mathrm{const}) \; .
\lab{F-r0-const}
\er

$\phantom{aa}$(i) $AdS_2 \times S^2$ with constant vacuum electric field
$|F_{0\eta}|\equiv |\vec{E}| = |c_F|$: 
\be
A(\eta) = 4\pi \llb c_F^2 - \sqrt{2}f_0 |c_F| - \frac{\L_0}{4\pi}\rrb\, \eta^2 
\quad (\eta - \mathrm{Poincare ~patch ~coordinate}) \; ,
\lab{AdS2}
\ee
provided either
$|c_F| > \frac{f_0}{\sqrt{2}}\Bigl( 1 + \sqrt{1 + \frac{\L_0}{2\pi f_0^2}}\Bigr)$
for $\L_0 \geq - 2\pi f_0^2$ or $|c_F| > \sqrt{\frac{1}{4\pi}|\L_0|}$ for
$\L_0 < 0 \; ,\; |\L_0| > 2\pi f_0^2$.

$\phantom{aa}$(ii) $Rind_2 \times S^2$ with constant vacuum electric field 
$|F_{0\eta}| = |c_F|$, where $Rind_2$ is the flat 2-dimensional Rindler spacetime with:
\be
A(\eta) = \eta \;\; \mathrm{for}\; 0 < \eta < \infty \quad \mathrm{or} \quad
A(\eta) = - \eta \;\; \mathrm{for}\; -\infty <\eta < 0 
\lab{Rindler2}
\ee
provided 
$|c_F| = \frac{f_0}{\sqrt{2}}\Bigl( 1 + \sqrt{1 + \frac{\L_0}{2\pi f_0^2}}\Bigr)$
for $\L_0 > - 2\pi f_0^2$.

$\phantom{aa}$(iii)  $dS_2 \times S^2$ with weak const vacuum electric field
$|F_{0\eta}| = |c_F|$, where $dS_2$ is the 2-dimensional de Sitter space with:
\be
A(\eta) = 1 - 4\pi\llb\sqrt{2}f_0 |c_F| -c_F^2 +\frac{\L_0}{4\pi}\rrb)\,\eta^2 \; , 
\lab{dS2}
\ee
when $|c_F| < \frac{f_0}{\sqrt{2}}\Bigl( 1 + \sqrt{1 + \frac{\L_0}{2\pi f_0^2}}\Bigr)$ 
for $\L_0 > - 2\pi f_0^2$. Note that $dS_2$ has {\em two horizons} at
$\eta = \pm \eta_0 \equiv 
\pm \Bigl\lb 4\pi\(\sqrt{2}f_0|c_F| - c_F^2\) + \L_0 \Bigr\rb^{-\h}$.

\section{Bulk Gravity/Nonlinear Gauge Field Coupled to Lightlike Brane Sources}

In the following two Sections we will consider bulk Einstein/non-linear gauge field 
system \rf{gravity+GG} self-consistently coupled to $N\geq 1$ (distantly separated) 
charged codimension-one {\em lightlike} $p$-brane (\textsl{LL-brane}) sources 
(here $p=2$). 

World-volume \textsl{LL-brane} actions in a reparametrization-invariant
Nambu-Goto-type or in an equivalent Polyakov-type formulation were proposed in 
Refs.\ct{KerrWH-varna2008-rotWH-ERbridge-BRkink}:
\br
S_{\rm LL}\lb q\rb  = - \h \int d^{p+1}\s\, T b_0^{\frac{p-1}{2}}\sqrt{-\g}
\llb \g^{ab} {\bar g}_{ab} - b_0 (p-1)\rrb \; ,
\lab{LL-action+EM} \\
{\bar g}_{ab} \equiv \pa_a X^\m G_{\m\n} \pa_b X^\n 
- \frac{1}{T^2} (\pa_a u + q\cA_a)(\pa_b u  + q\cA_b) 
\;\; , \;\; \cA_a \equiv \pa_a X^\m A_\m \; .
\lab{ind-metric-ext-A}
\er
Here and below the following notations are used:
\begin{itemize}
\item
$\g_{ab}$ is the {\em intrinsic} world-volume Riemannian metric;\\
$g_{ab}=\pa_a X^{\m} G_{\m\n}(X) \pa_b X^{\n}$ is the {\em induced} metric on the 
world-volume, which becomes {\em singular} on-shell (manifestation of the lightlike 
nature); $b_0$ is world-volume ``cosmological constant''.
\item
$X^\m (\s)$ are the $p$-brane embedding coordinates in the bulk
$D$-dimensional spacetime with Riemannian metric
$G_{\m\n}(x)$ ($\m,\n = 0,1,\ldots ,D-1$); 
$(\s)\equiv \(\s^0 \equiv \t,\s^i\)$ with $i=1,\ldots ,p$ ;
$\pa_a \equiv \partder{}{\s^a}$.
\item
$u$ is auxiliary world-volume scalar field defining the lightlike direction
of the induced metric;
\item
$T$ is {\em dynamical (variable)} brane tension;
\item
$q$ -- the coupling to bulk spacetime gauge field $\cA_\m$ is
\textsl{LL-brane} surface charge density.
\end{itemize}

The on-shell singularity of the induced metric $g_{ab}$ , \textsl{i.e.}, 
the lightlike property, directly follows from the \textsl{LL-brane} equations of
motion:
\be
g_{ab} \({\bar g}^{bc}(\pa_c u  + q\cA_c)\) = 0 \; .
\lab{on-shell-singular-A}
\ee

Now, let us consider the full action of self-consistently coupled bulk 
Einstein/non-linear gauge field/\textsl{LL-brane} system
($L(F^2) = - \frac{1}{4} F^2 - \frac{f_0}{2} \sqrt{- F^2}$):
\be
S = \int d^4 x \sqrt{-G} \Bigl\lb \frac{R(G) - 2\L_0}{16\pi} + L(F^2)\Bigr\rb 
+ \sum_{k=1}^N S_{\mathrm{LL}}\lb q^{(k)}\rb \; ,
\lab{gravity+GG+LL}
\ee
where the superscript $(k)$ indicates the $k$-th \textsl{LL-brane}.

The corresponding equations of motion are as follows:
\br
R_{\m\n} - \h G_{\m\n} R + \L_0 G_{\m\n} = 
8\pi \Bigl\lb T^{(F)}_{\m\n} + \sum_{k=1}^N T^{(k)}_{\m\n}\Bigr\rb \; ,
\lab{einstein+LL-eqs} \\
\pa_\n \Bigl\lb\sqrt{-G} \Bigl( 1 - \frac{f_0}{\sqrt{-F^2}}\Bigr) 
F_{\k\l} G^{\m\k} G^{\n\l}\Bigr\rb + \sum_{k=1}^N j_{(k)}^\m = 0 \; .
\lab{GG+LL-eqs}
\er

The energy-momentum tensor and the charge current density of $k$-th 
\textsl{LL-brane} are straightforwardly derived from the pertinent \textsl{LL-brane} 
world-volume action \rf{LL-action+EM}:
\be
T_{(k)}^{\m\n} = 
- \int\!\! d^3\s\,\frac{\d^{(4)}\Bigl(x-X_{(k)}(\s)\Bigr)}{\sqrt{-G}}
\, T^{(k)}\,\sqrt{|{\bar g}_{(k)}|} {\bar g}_{(k)}^{ab}
\pa_a X_{(k)}^\m \pa_b X_{(k)}^\n \; ,
\lab{T-brane-A}
\ee
\be
j_{(k)}^\m = -
q^{(k)} \int\!\! d^3\s\,\d^{(4)}\Bigl(x-X_{(k)}(\s)\Bigr)
\sqrt{|{\bar g}_{(k)}|} {\bar g}_{(k)}^{ab}\pa_a X_{(k)}^\m 
\frac{\pa_b u^{(k)} + q^{(k)}\cA^{(k)}_b}{T^{(k)}} \; .
\lab{j-brane-A}
\ee

Solving Eqs.\rf{einstein+LL-eqs}--\rf{GG+LL-eqs} with \rf{T-brane-A}--\rf{j-brane-A}
we find ``thin-shell'' wormhole solutions of static ``spherically-symmetric'' type 
(in Eddington-Finkelstein coordinates 
$dt=dv-\frac{d\eta}{A(\eta)}\; ,\; F_{0\eta} = F_{v\eta}$):
\br
ds^2 = - A(\eta) dv^2 + 2dv d\eta + C(\eta) h_{ij}(\th) d\th^i d\th^j \quad ,\quad
F_{v\eta} = F_{v\eta} (\eta)\; , 
\lab{static-spherical-EF} \\
-\infty < \eta < \infty \quad, \;\; A(\eta^{(k)}_0) = 0 \;\; 
\mathrm{for} \;\; \eta^{(1)}_0 <\ldots<\eta^{(N)}_0 \; .
\lab{common-horizons}
\er
The derivation of these ``thin-shell'' wormhole solutions proceeds along the
following main steps:

(i) Take ``vacuum'' solutions of \rf{einstein+LL-eqs}--\rf{GG+LL-eqs}
(without delta-function \textsl{LL-brane} terms) in each spacetime region
(separate ``universe'') given by 
$\bigl(-\infty\! <\!\eta\!<\!\eta^{(1)}_0\bigr),\ldots,$
$\bigl(\eta^{(N)}_0 \!<\!\eta \!<\!\infty\bigr)$ with common horizon(s) at 
$\eta=\eta^{(k)}_0$ ($k=1,\ldots ,N$).

(ii) Each $k$-th \textsl{LL-brane} automatically locates itself on the
horizon at $\eta=\eta^{(k)}_0$ -- intrinsic property of \textsl{LL-brane} dynamics
\ct{KerrWH-varna2008-rotWH-ERbridge-BRkink}.

(iii) Match discontinuities of the derivatives of the metric and
the gauge field strength  across each horizon at $\eta=\eta^{(k)}_0$ using the 
explicit expressions for the \textsl{LL-brane} 
stress-energy tensor and charge current density \rf{T-brane-A}--\rf{j-brane-A}.

\section{Charge ``Hiding''and Charge Confining Wormholes}

First we will construct ``one-throat'' wormhole solutions to \rf{gravity+GG+LL} with
the charged \textsl{LL-brane} occupying the wormhole ``throat'', which connects
(i) a non-compact ``universe'' with Reissner-Nordstr{\"o}m-(anti)-de-Sitter-type
geometry (where the cosmological constant is 
partially or entirely {\em dynamically} generated) to (ii) a compactified 
(``tubelike'') ``universe'' of (generalized) Levi-Civita-Bertotti-Robinson type 
with geometry $AdS_2 \times S^2$ or $Rind_2 \times S^2$. 

These wormholes possess the novel property of {\em hiding} electric charge from 
external observer in the non-compact ``universe''. Namely, the whole electric flux 
produced by the charged \textsl{LL-brane} at the wormhole ``throat'' is pushed into 
the ``tubelike'' ``universe''. As a result, the non-compact ``universe'' becomes
electrically neutral with Schwarzschild-(anti-)de-Sitter or purely Schwarzschild
geometry. Therefore, an external observer in the non-compact ``universe''
detects a {\em genuinely charged} matter source (the charged \textsl{LL-brane}) 
as {\em electrically neutral}. 

The explicit form
$ds^2 = - A(\eta) dv^2 + 2dv d\eta + C(\eta) \( d\th^2 + \sin^2 \th d\vp^2\)$
for the metric and the nonlinear gauge theory's electric field $F_{v\eta}(\eta)$ read:
\begin{itemize}
\item
``Left universe'' of Levi-Civita-Bertotti-Robinson (``tubelike'') type 
with geometry $AdS_2 \times S^2$ for $\eta< 0$:
\br
A(\eta) = 4\pi \( c_F^2 - \sqrt{2}f_0|c_F| - \frac{\L_0}{4\pi}\)\,\eta^2
\; ,\; C(\eta) \equiv r_0^2 = \frac{1}{4\pi c_F^2 + \L_0} \; ,
\lab{BR-AdS2-left-a} \\
|F_{v\eta}| \equiv |\vec{E}| = |c_F| > 
\frac{f}{\sqrt{2}}\Bigl( 1+\sqrt{1+\frac{\L_0}{2\pi f_0^2}}\,\Bigr)
\quad \mathrm{for}\; \L_0 > - 2\pi f_0^2 \; ,
\nonu\\
\mathrm{or}\quad |F_{v\eta}|\equiv |\vec{E}| = |c_F| > \sqrt{\frac{1}{4\pi}|\L_0|}
\quad \mathrm{for}\; \L_0 <0 \;,\; |\L_0| > 2\pi f_0^2 \; .
\nonu
\er
\item
Non-compact ``right universe'' for $\eta> 0$ comprising the exterior region of 
Reissner-Nordstr{\"o}m-de-Sitter-type black hole beyond the middle
(Schwarzschild-type) horizon $r_0$ when $\L_0 > - 2\pi f_0^2$ (in particular, when
$\L_0 = 0$), or the exterior region of 
Reissner-Nordstr{\"o}m-{\em anti}-de-Sitter-type black hole beyond the outer 
(Schwarzschild-type) horizon $r_0$ in the case $\L_0 <0$ and $|\L_0| > 2\pi f_0^2$,
or the exterior region of Reissner-Nordstr{\"o}m-``hedgehog'' black hole for 
$|\L_0| = 2\pi f_0^2$ (note: $A(\eta) \equiv A_{\mathrm{RN-((A)dS)}} (r_0 + \eta)$):
\br
A(\eta) = 1 - \sqrt{8\pi}|Q|f_0 - \frac{2m}{r_0 + \eta} + 
\frac{Q^2}{(r_0 + \eta)^2} - \frac{\L_0 + 2\pi f_0^2}{3} (r_0 + \eta)^2 \; ,
\nonu \\
\phantom{aaa}
\lab{RNdS-right-a} \\
C(\eta) = (r_0 + \eta)^2 \quad ,\quad 
|F_{v\eta}| \equiv |\vec{E}| = 
\frac{f_0}{\sqrt{2}} + \frac{|Q|}{\sqrt{4\pi}\, (r_0 + \eta)^2} \; .
\nonu
\er
\end{itemize}

The matching relations for the discontinuities of the metric and gauge field
components across the \textsl{LL-brane} world-volume occupying the
wormhole ``throat'' (which are here derived self-consistently from a
well-defined world-volume Lagrangian action principle for the \textsl{LL-brane}) 
\rf{LL-action+EM} determine all parameters of the wormhole solutions as
functions of $q$ (the \textsl{LL-brane} charge) and $f_0$ (coupling constant of 
$\sqrt{-F^2}$):
\be
Q=0 \quad ,\quad |c_F| = |q| + \frac{f_0}{\sqrt{2}} \; ,
\lab{param-1}
\ee
as well as the allowed range for the ``bare'' cosmological constant: 
\be
-4\pi\Bigl(|q|+\frac{f_0}{\sqrt{2}}\Bigr)^2 < 
\L_0 < 4\pi\Bigl(q^2 -\frac{f_0^2}{2}\Bigr) \; ,
\lab{CC-interval}
\ee
The relations \rf{param-1} 
(recall $|F_{v\eta}|\equiv |\vec{E}| = |c_F|$ in the ``tubelike'' ``left
universe'') have profound consequences:

(A) The non-compact ``right universe'' \rf{RNdS-right-a}
becomes exterior region of 
electrically neutral Schwarzschild-({\em anti}-)de-Sitter or purely
Schwarzschild black hole beyond the Schwarzschild horizon carrying a vacuum 
constant radial electric field $|F_{v\eta}|\equiv |\vec{E}|=\frac{f_0}{\sqrt{2}}$.

(B) Recalling that the dielectric displacement field is
$\vec{D} = \Bigl( 1 - \frac{f_0}{\sqrt{2}|\vec{E}|}\Bigr)\,\vec{E}$, we find
from the second relation \rf{param-1} that the whole flux produced by the charged
\textsl{LL-brane} flows only into the ``tubelike'' ``left universe'' 
\rf{BR-AdS2-left-a} (since
$\vec{D} = 0$ in the non-compact ``right universe''). This is a novel property of 
{\em hiding electric charge through a wormhole} connecting non-compact to a
``tubelike'' universe from external observer in the non-compact ``universe''.

The {\em charge-``hiding''} wormhole geometry is visualized on Fig.1 below.

\vspace{.1in}

\begin{figure}
\begin{center}
\includegraphics[scale=0.8,angle=270,keepaspectratio=true]{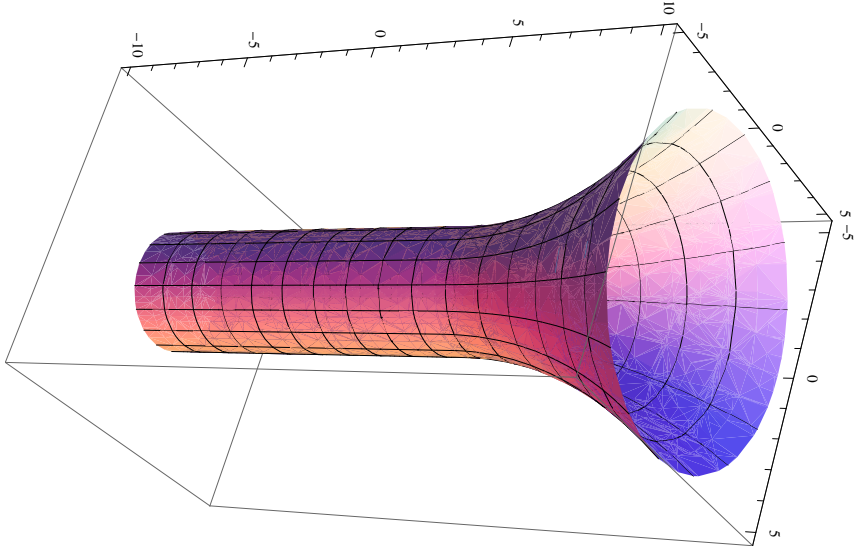}
\caption{Shape of $t=const$ and $\th=\frac{\pi}{2}$ slice of
{\em charge-``hiding''} wormhole geometry: the whole electric flux produced
by the charged \textsl{LL-brane} at the ``throat'' is expelled 
into the left infinitely long cylindric tube.}
\end{center}
\end{figure}

Further, we find  more interesting ``two-throat'' wormhole solution exhibiting 
{\em QCD-like charge confinement} effect -- obtained from a self-consistent 
coupling of the gravity/nonlinear-gauge-field system 
with two identical {\em oppositely charged} \textsl{LL-branes} 
(Eq.\rf{gravity+GG+LL} with $N=2$). The total ``two-throat''
wormhole spacetime manifold is made of:

(i) ``Left-most'' non-compact ``universe'' comprising the exterior region of 
Reissner-Nordstr{\"o}m-de-Sitter-type black hole beyond the middle Schwarzschild-type 
horizon $r_0$ for the ``radial-like'' $\eta$-coordinate interval:
\be
-\infty < \eta < -\eta_0 
\equiv - \Bigl\lb 4\pi\(\sqrt{2}f_0|c_F| - c_F^2\) + \L_0 \Bigr\rb^{-\h} \; ,
\lab{left-interval}
\ee
where:
\br
A(\eta) = A_{\mathrm{RNdS}}(r_0 - \eta_0 - \eta)= \phantom{aaaaaaaaaaaaaa}
\nonu\\
1 - \sqrt{8\pi}|Q|f_0 - \frac{2m}{r_0 - \eta_0 - \eta} + 
\frac{Q^2}{(r_0 - \eta_0 - \eta)^2} 
- \frac{\L_0 + 2\pi f_0^2}{3} (r_0 - \eta_0 - \eta)^2 \; ,
\nonu\\
\phantom{aaa}
\lab{RNdS-left-most-1}\\
C(\eta) = (r_0 - \eta_0 - \eta)^2 \;\; ,\;\;
|F_{v\eta}(\eta)| \equiv |\vec{E}| = \frac{f_0}{\sqrt{2}} 
+ \frac{|Q|}{\sqrt{4\pi}\,(r_0 - \eta_0 - \eta)^2} \; .\phantom{aaa}
\nonu
\er

(ii) ``Middle'' ``tube-like'' ``universe'' of
Levi-Civita-Bertotti-Robinson type with geometry $dS_2 \times S^2$ 
 comprising the finite extent (w.r.t. $\eta$-coordinate) 
region between the two horizons of $dS_2$ at $\eta = \pm \eta_0$:
\be
-\eta_0 < \eta < \eta_0 
\equiv \Bigl\lb 4\pi\(\sqrt{2}f_0|c_F| - c_F^2\) + \L_0 \Bigr\rb^{-\h} \; ,
\lab{middle-interval}
\ee
where the metric coefficients and electric field are:
\br
A(\eta) = 1 - \Bigl\lb 4\pi\(\sqrt{2}f_0|c_F| - c_F^2\) + \L_0 \Bigr\rb\,\eta^2
\;\; ,\;\; A(\pm \eta_0) = 0 \; ,
\nonu\\
\phantom{aaa}
\lab{LCBR-middle-1} \\
C(\eta) = r_0^2 = \frac{1}{4\pi c_F^2 + \L_0} \quad ,\quad
|F_{v\eta}|\equiv |\vec{E}| = |c_F| < 
\frac{f}{\sqrt{2}}\Bigl( 1 + \sqrt{1 + \frac{\L}{2\pi f_0^2}}\Bigr) \; ,
\nonu
\er
with $\L_0 > - 2\pi f_0^2$;

(iii) ``Right-most'' non-compact ``universe'' comprising the exterior region of 
Reissner-Nordstr{\"o}m-de-Sitter-type black hole beyond the middle Schwarzschild-type 
horizon $r_0$ for the ``radial-like'' $\eta$-coordinate interval
~$\eta_0 < \eta < \infty$ ($\eta_0$ as in \rf{middle-interval}), where:
\br
A(\eta) = A_{\mathrm{RNdS}}(r_0 + \eta - \eta_0)  \phantom{aaaaaaaaaa}
\nonu \\
= 1 - \sqrt{8\pi}|Q|f_0 - \frac{2m}{r_0 + \eta - \eta_0} + 
\frac{Q^2}{(r_0 + \eta - \eta_0)^2} 
- \frac{\L_0 + 2\pi f_0^2}{3} (r_0 + \eta - \eta_0)^2 \; ,
\nonu \\
\phantom{aaa}
\lab{RNdS-right-most-1} \\
C(\eta) = (r_0 + \eta - \eta_0)^2 \;\; ,\;\;
|F_{v\eta}(\eta)| \equiv |\vec{E}| = \frac{f_0}{\sqrt{2}} 
+ \frac{|Q|}{\sqrt{4\pi}\,(r_0 + \eta - \eta_0)^2} \; .
\nonu
\er
As dictated by the \textsl{LL-brane} dynamics 
\ct{KerrWH-varna2008-rotWH-ERbridge-BRkink} each of the two \textsl{LL-branes}
locates itself on one of the two common horizons at $\eta = \pm \eta_0$
between ``left'' and ``middle'', and between ``middle'' and ``right''
``universes'', respectively.

The matching relations for the discontinuities of the metric and gauge field
components across the each of the two \textsl{LL-brane} world-volumes
determine all parameters of the wormhole solutions as functions of $\pm q$
(the opposite \textsl{LL-brane} charges) and $f_0$ (coupling constant of 
$\sqrt{-F^2}$). Most importantly we obtain:
\be
Q=0 \quad ,\quad |c_F| = |q| + \frac{f_0}{\sqrt{2}} \; ,
\lab{param-1-conf}
\ee
and the bare cosmological constant must be in the interval:
\be
\L_0 \leq 0 \quad ,\quad |\L_0| < 2\pi (f_0^2 - 2 q^2) 
\quad \to \quad |q| < \frac{f_0}{\sqrt{2}} \; ,
\lab{CC-interval-conf}
\ee
in particular, $\L_0$ could be zero. 

Similarly to the charge-``hiding'' case, relations \rf{param-1-conf} meaning:
\br
|\vec{E}|_{\rm middle ~universe} = |q| + |\vec{E}|_{\rm left/right ~universe}
\; ,
\nonu
\er
have profound consequences:
\begin{itemize}
\item
The ``left-most'' \rf{RNdS-left-most-1} and ``right-most'' \rf{RNdS-right-most-1}
non-compact ``universes'' become
two identical copies of the {\em electrically neutral} exterior region of
Schwarzschild-de-Sitter black hole beyond the Schwarzschild horizon. They both
carry a constant vacuum radial electric field with magnitude 
$|\vec{E}|=\frac{f_0}{\sqrt{2}}$ pointing inbound towards the
horizon in one of these ``universes'' and pointing outbound w.r.t. the horizon
in the second ``universe''. The corresponding electric displacement field 
$\vec{D}=0$, so there is {\em no} electric flux there (recall
$\vec{D} = \Bigl( 1 - \frac{f_0}{\sqrt{2}|\vec{E}|}\Bigr)\,\vec{E}$).
\item
The whole electric flux produced by the two charged \textsl{LL-branes} with 
opposite charges $\pm q$ at the boundaries of the above non-compact ``universes''
is {\em confined} within the ``tube-like'' middle ``universe'' \rf{LCBR-middle-1}
of Levi-Civita-Robinson-Bertotti type
with geometry $dS_2 \times S^2$, where the constant electric field is
$|\vec{E}|=\frac{f_0}{\sqrt{2}} + |q|$ with associated non-zero electric 
displacement field $|\vec{D}|= |q|$ . This is {\em QCD-like confinement}.
\end{itemize}

A simple visualization of the {\em charge-confining} wormhole geometry is given in 
Fig.2.

\begin{figure}
\begin{center}
\includegraphics[scale=0.8,angle=270,keepaspectratio=true]{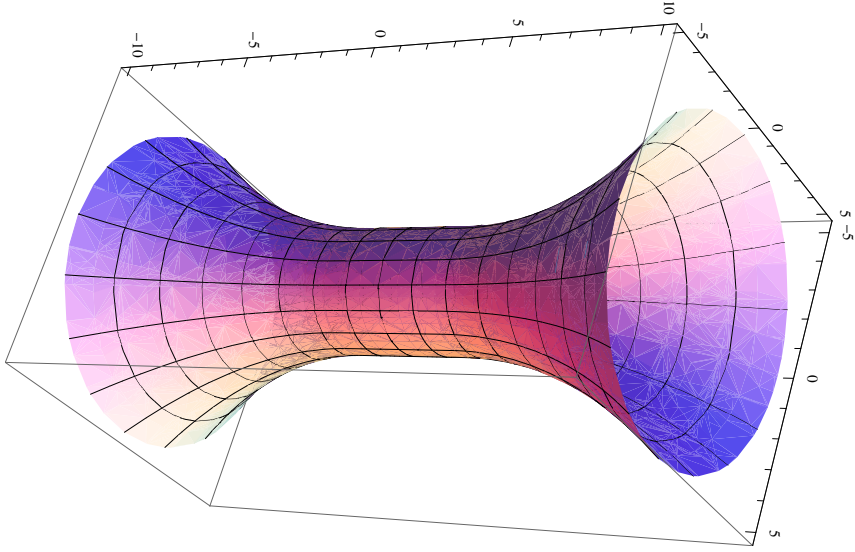}
\caption{Shape of $t=const$ and $\th=\frac{\pi}{2}$ slice of
{\em charge-confining} wormhole geometry: the whole electric flux produced
by the two oppositely charged \textsl{LL-branes} is confined
within the middle finite-extent cylindric tube between the ``throats''.}
\end{center}
\end{figure}

\section{$R^2$-Gravity Coupled to Confining Nonlinear Gauge Field and Dilaton}

Consider now coupling of $f(R)= R + \a R^2$ gravity (possibly
with a bare cosmological constant $\L_0$) to a ``dilaton'' $\phi$ and
the nonlinear gauge field system containing $\sqrt{-F^2}$:
\br
S = \int d^4 x \sqrt{-g} \Bigl\lb \frac{1}{16\pi} 
\Bigl( f\bigl(R(g,\G)\bigr) - 2\L_0 \Bigr) + L(F^2(g)) + L_D (\phi,g) \Bigr\rb \; ,
\lab{f-gravity+GG+D} \\
f\bigl(R(g,\G)\bigr) = R(g,\G) + \a R^2(g,\G) \quad ,\quad 
R(g,\G) = R_{\m\n}(\G) g^{\m\n} \; ,
\lab{f-gravity} \\
L(F^2(g)) = - \frac{1}{4e^2} F^2(g) - \frac{f_0}{2} \sqrt{- F^2(g)} \; ,
\lab{GG-g} \\
F^2(g) \equiv F_{\k\l} F_{\m\n} g^{\k\m} g^{\l\n} \;\; ,\;\;
F_{\m\n} = \pa_\m A_\n - \pa_\n A_\m \;
\lab{F2-g} \\
L_D (\phi,g) = -\h g^{\m\n}\pa_\m \phi \pa_\n \phi - V(\phi) \; .
\lab{L-dilaton}
\er
$R_{\m\n}(\G)$ is the Ricci curvature in the first order (Palatini) formalism, 
\textsl{i.e.}, the spacetime metric $g_{\m\n}$ and the affine connection 
$\G^\m_{\n\l}$ are \textsl{a priori} independent variables.

The equations of motion resulting from the action \rf{f-gravity+GG+D} read:
\be
R_{\m\n}(\G) = \frac{1}{f^{\pr}_R}\llb 8\pi T_{\m\n} + 
\h f\bigl(R(g,\G)\bigr) g_{\m\n}\rrb 
\;\; ,\;\; f^{\pr}_R \equiv \frac{df(R)}{dR} = 1 + 2\a R(g,\G) \; ,
\lab{g-eqs}
\ee
\br
\nabla_\l \(\sqrt{-g} f^{\pr}_R g^{\m\n}\)  = 0 \; ,
\lab{gamma-eqs} \\
\pa_\n \Bigl(\sqrt{-g} \Bigl\lb 1/e^2 - \frac{f_0}{\sqrt{-F^2(g)}}
\Bigr\rb F_{\k\l} g^{\m\k} g^{\n\l}\Bigr)=0 \; .
\lab{GG-eqs-R2}
\er
The total energy-momentum tensor is given by:
\br
T_{\m\n} = 
\Bigl\lb L(F^2(g))+L_D (\phi,g)-\frac{1}{8\pi}\L_0\Bigr\rb g_{\m\n} 
\nonu \\
+ \Bigl(1/e^2 - \frac{f_0}{\sqrt{-F^2(g)}}\Bigr) 
F_{\m\k} F_{\n\l} g^{\k\l} + \pa_\m \phi \pa_\n \phi \;.
\lab{T-total}
\er
Eq.\rf{gamma-eqs} leads to the relation $\nabla_\l \( f^{\pr}_R g_{\m\n}\)=0$
and thus it implies transition to the ``physical'' Einstein-frame metrics 
$h_{\m\n}$ via conformal rescaling of the original metric $g_{\m\n}$ \ct{olmo-etal}:
\be
g_{\m\n} = \frac{1}{f^{\pr}_R} h_{\m\n} \quad ,\quad
\G^\m_{\n\l} = \h h^{\m\k} \(\pa_\n h_{\l\k} + \pa_\l h_{\n\k} - \pa_\k h_{\n\l}\)
\; .
\lab{einstein-frame}
\ee
Using \rf{einstein-frame} the $R^2$-gravity equations of motion \rf{g-eqs} can be
rewritten in the form of {\em standard} Einstein equations:
\be
R^\m_\n (h) = 8\pi \({T_{\rm eff}}^\m_\n (h) - \h \d^\m_\n {T_{\rm eff}}^\l_\l (h)\)
\lab{einstein-h-eqs}
\ee
with effective energy-momentum tensor of the following form:
\be
{T_{\rm eff}}_{\m\n} (h) = h_{\m\n} L_{\rm eff} (h) 
- 2 \partder{L_{\rm eff}}{h^{\m\n}} \; .
\lab{T-h-eff}
\ee
The effective Einstein-frame matter Lagrangian reads (the dilaton kinetic term
$X(\phi,h) \equiv -\h h^{\m\n}\pa_\m \phi \pa_n \phi$ will be ignored
in the sequel):
\br
L_{\rm eff} (h) = - \frac{1}{4 e_{\rm eff}^2 (\phi)} F^2(h) 
- \h f_{\rm eff} (\phi) \sqrt{- F^2(h)} 
\nonu \\
+ \frac{X(\phi,h)\bigl(1+16\pi\a X(\phi,h)\bigr) - V(\phi) -\L_0/8\pi
}{1+8\a\( 8\pi V(\phi)+\L_0\)}
\lab{L-eff-h}
\er
with the following dynamical $\phi$-dependent couplings:
\br
\frac{1}{e_{\rm eff}^2 (\phi)} = \frac{1}{e^2} + 
\frac{16\pi\a f_0^2}{1 + 8\a \(8\pi V(\phi) + \L_0 \)} \; ,
\lab{e-eff} \\
f_{\rm eff}(\phi)=f_0 \frac{1+32\pi\a X(\phi,h)}{1 + 8\a\(8\pi V(\phi)+\L_0\)}
\; .
\lab{f-eff}
\er

Thus, all equations of motion of the original $R^2$-gravity system 
\rf{f-gravity+GG+D}--\rf{L-dilaton} can be equivalently derived from the 
following Einstein/nonlinear-gauge-field/dilaton action:
\be
S_{\rm eff} = \int d^4 x \sqrt{-h} \Bigl\lb \frac{R(h)}{16\pi} 
+ L_{\rm eff} (h)\Bigr\rb \; ,
\lab{einstein-frame-action}
\ee
where $R(h)$ is the standard Ricci scalar of the metric $h_{\m\n}$ and 
$L_{\rm eff} (h)$ is as in \rf{L-eff-h}.

\textbf{Important observation}. Even if ordinary kinetic Maxwell term
$-\frac{1}{4}F^2$ is absent in the original system ($e^2 \to \infty$ in \rf{GG-g}),
such term is nevertheless {\em dynamically generated} in the Einstein-frame action 
\rf{L-eff-h}--\rf{einstein-frame-action}, which is a {\em combined effect} of
$\a R^2$ and $-\frac{f_0}{2}\sqrt{-F^2}$:
\be
S_{\rm maxwell} =
-4\pi\a f_0^2 \int d^4 x \sqrt{-h} \frac{F_{\k\l} F_{\m\n} h^{\k\m} h^{\l\n}}{
1+8\a\(8\pi V(\phi)+\L_0\)} \; .
\lab{dynamical-maxwell}
\ee

In what follows we consider constant ``dilaton'' $\phi$ extremizing the effective
Lagrangian \rf{L-eff-h}:
\br
L_{\rm eff} =
- \frac{1}{4 e_{\rm eff}^2 (\phi)} F^2(h) - \h f_{\rm eff} (\phi) \sqrt{-F^2(h)}
- V_{\rm eff}(\phi) \; ,
\lab{L-eff-0} \\
V_{\rm eff}(\phi) = \frac{V(\phi) + \frac{\L_0}{8\pi}}{1+8\a\(8\pi V(\phi)+\L_0\)}
\;\; ,\;\; f_{\rm eff} (\phi) = \frac{f_0}{1+8\a\(8\pi V(\phi)+\L_0\)} \; ,
\lab{V-f-eff-1} \\
\frac{1}{e_{\rm eff}^2 (\phi)} = 
\frac{1}{e^2} + \frac{16\pi\a f_0^2}{1 + 8\a \(8\pi V(\phi) + \L_0 \)} \; .
\lab{e-eff-1}
\er
\textbf{Important observation}. 
The dynamical couplings and effective potential are extremized 
{\em simultaneously} -- this is an explicit realization of
``least coupling principle'' of Damour-Polyakov \ct{damour-polyakov}:
\be
\partder{f_{\rm eff}}{\phi} = - 64\pi\a f_0 \partder{V_{\rm eff}}{\phi}
\;\; ,\;\; \partder{}{\phi}\frac{1}{e_{\rm eff}^2} =
-(32\pi\a f_0)^2 \partder{V_{\rm eff}}{\phi} 
\;\; \to \partder{L_{\rm eff}}{\phi} \sim \partder{V_{\rm eff}}{\phi} \; .
\lab{f-e-extremize}
\ee
Therefore at the extremum of $L_{\rm eff}$ \rf{L-eff-0} $\phi$ must satisfy:
\be
\partder{V_{\rm eff}}{\phi} = 
\frac{V^{\pr}(\phi)}{\llb 1+8\a\(\k^2 V(\phi)+\L_0\)\rrb^2} = 0 \; .
\lab{V-extremum}
\ee
There are two generic cases:

$\phantom{aa}$(a) {\em Confining phase}: Eq.\rf{V-extremum} is satisfied for some 
finite-value $\phi_0$ extremizing the original potential $V(\phi)$: 
$V^{\pr}(\phi_0) = 0$.

$\phantom{aa}$(b) {\em Deconfinement phase}: For polynomial or exponentially 
growing original $V(\phi)$, so that $V(\phi) \to \infty$ when $\phi \to \infty$, 
we have:
\be
\partder{V_{\rm eff}}{\phi} \to 0 \quad ,\quad 
V_{\rm eff} (\phi) \to \frac{1}{64\pi\a} = {\rm const} \quad {\rm when} \;\;
\phi \to \infty \; ,
\lab{flat-region}
\ee
\textsl{i.e.}, for sufficiently large values of $\phi$ we find a ``flat region''
in $V_{\rm eff}$. This ``flat region'' triggers a {\em transition from 
confining to deconfinement dynamics}.

Namely, in the ``flat-region'' case ($V(\phi) \to \infty$) we have from 
\rf{V-f-eff-1}--\rf{e-eff-1}:
\be
f_{\rm eff} \to 0 \quad ,\quad e^2_{\rm eff} \to e^2
\lab{deconfine}
\ee
and the effective gauge field Lagrangian \rf{L-eff-0} reduces to the ordinary
\textsl{non-confining} one (the ``square-root'' term $\sqrt{-F^2}$ vanishes):
\be
L^{(0)}_{\rm eff} = -\frac{1}{4e^2} F^2(h) - \frac{1}{64\pi\a}
\lab{L-eff-h-0}
\ee
with an {\em induced} cosmological constant $\L_{\rm eff} = 1/8\a$, which is
{\em completely independent} of the bare cosmological constant $\L_0$.

Within the physical ``Einstein''-frame in the confining phase 
($V^{\pr}(\phi_0) = 0 \; ,\, \phi_0 =\mathrm{finite}$) we find:

(A) Reissner-Nordstr{\"o}m-({\em anti}-)de-Sitter type black holes, in
particular,  non-standard Reissner-Nordstr{\"o}m type with non-flat ``hedgehog''
asymptotics, generalizing solutions \rf{spherical-static}--\rf{cornell-sol}
in the ordinary Einstein-gravity case, where now the effective cosmological constant
and the vacuum constant radial electric field read:
\br
\L_{\rm eff}(\phi_0) = \frac{\L_0 +8\pi V(\phi_0)+2\pi e^2 f^2_0}{
1+8\a\(\L_0 +8\pi V(\phi_0)+2\pi e^2 f^2_0\)} \; ,
\lab{h-CC-eff}\\
|\vec{E}_{\rm vac}| = \Bigl(\frac{1}{e^2} + 
\frac{16\pi\a f_0^2}{1 + 8\a\(8\pi V(\phi_0) + \L_0 \)}\Bigr)^{-1}
\frac{f_0/\sqrt{2}}{1+8\a\(8\pi V(\phi_0)+\L_0\)} \; .
\lab{vacuum-radial}
\er

(B) Levi-Civita-Bertotti-Robinson type ``tubelike'' spacetimes 
with geometries $AdS_2 \times S^2$, $Rind_2 \times S^2$ and $dS_2 \times S^2$
generalizing \rf{gen-BR-metric}--\rf{dS2},
where now (using short-hand notation $\L(\phi_0)\equiv 8\pi V(\phi_0) + \L_0$):
\be
\frac{1}{r_0^2} = \frac{4\pi}{1+8\a\L(\phi_0)}\Bigl\lb
\Bigl(1+8\a\(\L(\phi_0)+2\pi f_0^2\)\Bigr) \vec{E}^2 +
\frac{1}{4\pi}\L(\phi_0)\Bigr\rb \; .
\lab{r0-eq}
\ee

\section{Discussion}

Inclusion of the non-standard nonlinear ``square-root'' gauge field term
provides explicit realization of the old ``classic'' idea of `t Hooft \ct{tHooft}
about the nature of low-energy confinement dynamics. 
Coupling of nonlinear gauge theory containing $\sqrt{-F^2}$ to
gravity (Einstein or $f(R)=R + \a R^2$ plus scalar ``dilaton'') leads to a 
variety of remarkable effects:
\begin{itemize}
\item
Dynamical effective gauge couplings and dynamical induced cosmological
constant;
\item
New non-standard black hole solutions of Reissner-Nordstr{\"o}m-({\em anti}-)de-Sitter 
type carrying an additional constant vacuum electric field, in particular, 
non-standard Reissner-Nordstr{\"o}m type black holes with asymptotically non-flat 
``hedgehog'' \ct{hedgehog} behavior;
\item
``Cornell''-type \ct{cornell-potential} confining potential in charged test particle
dynamics;
\item
Coupling to a charged lightlike brane produces a charge-``hiding'' wormhole,
where a genuinely charged matter source is detected as electrically
neutral by an external observer;
\item
Coupling to two oppositely charged lightlike brane sources produces a 
two-``throat'' wormhole displaying a genuine QCD-like charge confinement.
\item
When coupled to $f(R)=R + \a R^2$ gravity plus scalar ``dilaton'', the
$\sqrt{-F^2}$ term triggers a transition from confining to deconfinement phase.
Standard Maxwell kinetic term for the gauge field is dynamically generated
even when absent in the original ``bare'' theory. 
The above are cumulative effects produced
by the {\em simultaneous} presence of $\a R^2$ and $\sqrt{-F^2}$ terms.
\end{itemize}

Let us conclude with a brief remark concerning the thermodynamic properties
of the non-standard black hole solutions described above. To this end, let us
recall that for any static spherically symmetric metric of the form
\rf{spherical-static} with Schwarzschild-type horizon $r_0$, \textsl{i.e.}, 
$A(r_0)=0\; ,\; \pa_r A\!\!\bv_{r_0} >0$, the so called {\em surface gravity} $\k$ 
proportional to Hawking temperature $T_h$ (\textsl{e.g.} \ct{BH-thermo}, Ch. 12.5) 
is given by $\k = 2\pi T_h = \h \pa_r A\!\!\bv_{r_0}$. With $A(r)$ of the general form
$A(r) = 1 - c(Q_i) - 2m/r + A_1 (r;Q_i)$, where $Q_i$ are the rest of the
black hole parameters apart from the mass $m$, and $c(Q_i)$ is generically a non-zero
constant as in \rf{CC-eff} (responsible for the ``hedgehog'' non-flat spacetime
asymptotics), one can straightforwardly derive the first law of black hole
thermodynamics for the above class of solutions:
\be
\d m = \frac{1}{8\pi}\k \d A_H + {\wti\P}_i \d Q_i \quad ,\; A_H = 4\pi r_0^2
\; ,\; {\wti\P}_i =
\frac{r_0}{2} \partder{}{Q_i} \Bigl( A_1 (r_0;Q_i) - c(Q_i)\Bigr) \; .
\lab{first-law}
\ee
In the special case of non-standard Reissner-Nordstr{\"o}m-(anti-)de-Sitter type 
black holes \rf{spherical-static}--\rf{CC-eff} with parameters $(m,Q)$ the
conjugate potential in \rf{first-law}:
\be
{\wti\P} = \sqrt{4\pi} \Bigl(\frac{Q}{\sqrt{4\pi} r_0}
- \frac{f_0}{\sqrt{2}}r_0\Bigr) =  \sqrt{4\pi} A_0 \bv_{r=r_0}
\lab{conjugate-potential}
\ee
is (up to a constant factor) the electric field potential of the nonlinear
gauge system on the horizon.

{\small
\section*{Acknowledgments}
E.N. is sincerely grateful to Prof. Branko Dragovich and the organizers
of the Seventh Meeting in Modern Mathematical Physics (Belgrade, Sept 2012) 
for cordial hospitality.
E.N. and S.P. are supported in part by Bulgarian NSF grant \textsl{DO 02-257}.
Also, all of us acknowledge support of our collaboration through the exchange
agreement between the Ben-Gurion University of the Negev and the Bulgarian Academy 
of Sciences.
}

\end{document}